# MULTI-CENTER CLINICAL TRIALS: RANDOMIZATION AND ANCILLARY STATISTICS

By Lu Zheng and Marvin Zelen

*Harvard School of Public Health*

The purpose of this paper is to investigate and develop methods for analysis of multi-center randomized clinical trials which only rely on the randomization process as a basis of inference. Our motivation is prompted by the fact that most current statistical procedures used in the analysis of randomized multi-center studies are model based. The randomization feature of the trials is usually ignored. An important characteristic of model based analysis is that it is straightforward to model covariates. Nevertheless, in nearly all model based analyses, the effects due to different centers and, in general, the design of the clinical trials are ignored. An alternative to a model based analysis is to have analyses guided by the design of the trial. Our development of design based methods allows the incorporation of centers as well as other features of the trial design. The methods make use of conditioning on the ancillary statistics in the sample space generated by the randomization process. We have investigated the power of the methods and have found that, in the presence of center variation, there is a significant increase in power. The methods have been extended to group sequential trials with similar increases in power.

**1. Introduction.** The randomized multi-center clinical trial is widely recognized as the ideal way to generate data to evaluate the benefit of therapies for the treatment of disease. The randomization process tends to eliminate bias introduced by physicians or patient preferences. In addition to randomization, sometimes double blind trials are used to eliminate such bias when the outcomes are subjective. Randomization also serves to balance unknown factors over treatments which may affect outcome. Randomization was first introduced by Fisher in 1935 [see Fisher (1971)] and was motivated by experimental design problems in agriculture. Randomization has proved to be equally important in studies on humans so that today randomized clinical trials are regarded as part of the foundation of the scientific basis of modern medicine.

Current practice in the analyses of randomized clinical trials is to use statistical methods which are model based; for example, linear, logistic and proportional hazard models. These methods are ideal for taking into account factors which influence outcome. However, the inference requires that subjects entering a trial constitute a random sample of subjects from a well defined population. Unfortunately this is rare. Subjects entering a trial are not a random sample of patients. We refer to these subjects as a "collection" of patients. Consequently, the basis of any inference is questionable in the absence of a random sample of subjects.

The randomization process can serve as a basis of inference and is an alternative to relying on a random sample as the basis of inference. In our view, in the absence of a random sample of patients, model based analyses are appropriate when they serve as approximations to a randomization analysis. However, relying on the randomization process limits the scope of the resulting inference. The inference strictly applies to the "collection" of subjects who have entered the trial. Any generalization of the inference to a population with disease must be made on how well the "collection" of patients in the trial approximates a well-defined disease population.

Under a population model, model-based methods test the null hypothesis of the equality of parameters from known distributions, while the null hypothesis of a randomization-based test is that the assignment of treatments A and B had no effect on the outcomes of the subjects enrolled in the study.

A 1988 special issue of Controlled Clinical Trials was devoted to discussing the statistical properties of randomization procedures in clinical trials. Special topics included simple randomization [Lachin (1988)], permuted block randomization [Matts and Lachin (1988)] and the urn-adaptive biased-coin randomization [Wei and Lachin (1988)]. Lachin, Matts, and Wei (1988), in the conclusion paper of the issue, shared the same view as ours. The main ideas in these papers have been summarized later in the book by Rosenberger and Lachin (2002).

The aim of this paper is to investigate randomization based analyses of randomized multi-center trials. A guiding principle in our development is that the analysis should take into account the design of the trial. Most randomized clinical trials are designed using permuted blocks. This feature of a trial is usually ignored in the model driven analyses. Dividing the patients into blocks enables balance among treatments with reference to a possible change in the population over time and the possibility of changing benefit over time; that is, physicians acquiring experience in the administration of the treatment will be better able to administer the treatment.

A widely accepted principle in frequentist inference is to condition the analysis on the ancillary statistics. Conditioning on ancillary statistics will reduce the sample space and generally will result in greater power compared to ignoring the ancillary statistics. Nevertheless, most frequentist analyses of randomized trials ignore the ancillary statistics. For example, ancillary statistics arise most naturally in randomized multi-center clinical trials in which the ancillary statistics are the number of patients assigned to each treatment within a center. Although the treatment sample sizes within each center are random variables, nearly all model-based analyses that incorporate center effects treat the sample sizes within centers as fixed quantities and ignore the probability aspects of the sample sizes. When the sample sizes within a center are large, this distinction may not be important. However in many multi-center trials there may be large numbers of centers which enter a relatively small number of patients. In such situations, modeling a parameter for each institution can result in a significant reduction in the precision of the test statistic. An alternative to adding more parameters is to reduce the sample space by conditioning [Pesarin (2001)]. The power of clinical trials may be increased by conditioning on the institutional sample sizes for each treatment. Conditioning on the sample sizes of centers is an illustration of how a randomization-based analysis of a trial may adjust for covariates. In this case the centers are the covariates. This idea generalizes when there are an arbitrary number of covariates.

Many approaches have been proposed to adjust for center effect in multi-center studies in the past. The majority of the adjustments use model-based methods, as reviewed by Localio, Berlin, Ten Have, and Kimmel (2001). The methods adjusting for center effects include the following: mixed-effects, random-effects, multistage hierarchical [Skene and Wakefield (1990), Matsuyama, Sakamoto, and Ohashi (1998)], Bayesian approach [Gray (1994), Yamaguchi, Ohashi, and Matsuyama (2002)] and frailty models for survival data [Andersen, Klein, and Zhang (1999)]. Boos and Brownie (1992) developed rank based methods to account for institution variation. Davis and Chung (1995) developed a Mantel–Haenszel mean score statistic using a randomization model for continuous or ordered categorical outcomes. The estimator, which is a weighted average of the center-specific mean differences, is equivalent to the estimate obtained using a fixed-effect model. Potthoff, Peterson, and George (2001) investigated several permutation tests for treatment-by-center interaction in multi-center clinical trials with survival outcomes. Brunner, Domhof, and Puri (2002) considered nonparametric tests where the statistics are weighted according to the different sample sizes within the levels of one factor. In general, there is very limited literature related to adjusting center effects using randomization-based methods.

The aim of this paper is to investigate randomization-based analyses, in which the experimental design is a permuted block design, and covariate adjustments are made by conditioning on the ancillary statistics. Comparisons are made with model-based analyses for linear, logistic and proportional hazard models. The conditioning on the ancillary statistics, in the presence of permuted blocks, generates some difficult combinatorial problems. To deal with this class of problems, we principally base our analyses on large sample procedures.

The paper is organized as follows. Section 2 formulates the problem and introduces notation. Section 3 presents the method for the analysis of multi-center trials using permuted block designs. Section 4 applies the idea to group sequential randomized clinical trials. Section 5 concludes with a discussion.

**2. Problem formulation and notation.** The key idea in using the randomization process as a basis for inference is that the only probability in a study is the introduction of uncertainty by the random assignment of treatments to patients. Patient outcomes are considered fixed and are not governed by probability distributions. The statistical procedures are based only on the randomization process and are distribution free. Conceptually, the distribution of the appropriate statistics can be obtained by enumerating the entire sample space. However, enumeration may not be feasible with large numbers of observations. Instead, the distribution may be approximated by sampling the sample space or using large sample approximations based on low order moments.

2.1. *Permuted blocks.* A major disadvantage of simple randomization is the possibility of generating unbalanced numbers of patients in treatments. Permuted block designs eliminate such possible imbalances. The basic idea of a permuted block design is to group patients into "blocks" according to the time entered in the study. Randomization is then carried out within each individual block, so that there is an equal number of subjects assigned to each treatment within a block. The application of permuted blocks is also viewed as a protection against unknown time trends in either the treatment effects or patient characteristics; that is, the application of the treatments may become more efficacious as more experience is gained with the treatments. A disadvantage of permuted block designs for single institution studies is that at certain allocations in the trial, a treatment assignment could be known to investigators, in advance of randomization. However, this phenomenon is unlikely in multi-center trials.

2.2. *Notation and permuted blocks*: *Single institution.* Consider a single-center randomized clinical trial comparing two treatments denoted by $A$ and $B$. Let a single permuted block contain $N$ patients. Suppose there are $P$ permuted blocks. Define the binary random variable

$$\delta_{ij} = \begin{cases} 1, & \text{if } i\text{th patient in } j\text{th block is assigned to } A, \\ 0, & \text{otherwise}, \end{cases}$$

where $i = 1, 2, \ldots, N$; $j = 1, 2, \ldots, P$. The treatments are randomly assigned, that is, $P\{\delta_{ij} = 1\} = \frac{1}{2}$ subject to $\sum_{i=1}^{N} \delta_{ij} = \frac{N}{2}$ for $j = 1, 2, \ldots, P$. The distribution properties of the $\{\delta_{ij}\}$ for the permuted blocks is well known and easily derived. The low order moments are

$$E\left[\delta_{ij}\bigg|\sum_{i=1}^{N}\delta_{ij} = \frac{N}{2}\right] = \frac{1}{2}, \quad \text{Var}\left(\delta_{ij}\bigg|\sum_{i=1}^{N}\delta_{ij} = \frac{N}{2}\right) = \frac{1}{4},$$

$$\text{Cov}\left(\delta_{ij}, \delta_{rj}\bigg|\sum_{i=1}^{N}\delta_{ij} = \frac{N}{2}\right) = -\frac{1}{4(N-1)} \quad \text{for } i \neq r.$$

Let $y_{ij}$ be the observed outcome for the $i$th patient in the $j$th block regardless of treatment assignment. This formulation assumes no difference between treatments. Then $S_A = \sum_{j=1}^{P}\sum_{i=1}^{N}\delta_{ij}y_{ij}$ and $S_B = \sum_{j=1}^{P}\sum_{i=1}^{N}(1-\delta_{ij})y_{ij}$ are the observed outcome totals for each treatment group. The outcomes $\{y_{ij}\}$ are assumed to be fixed quantities. Note that with this formulation, $S_A + S_B = S = \sum_{j=1}^{B}\sum_{i=1}^{N}y_{ij}$ is a fixed quantity.

A comparison of the two groups is usually carried out by comparing the distribution of the difference of the sample averages; that is,

$$D = \frac{S_A}{NP/2} - \frac{S_B}{NP/2} = \frac{2}{NP}(2S_A - S).$$

The inference on the treatment difference may then be made by considering the conditional randomization distribution of $D$ or, equivalently, of $S_A$, since $S_A$ is the only random variable in $D$. The randomization distribution of $S_A$ can be obtained by considering the randomization distribution in each block and taking the convolution among the blocks. For example, if $N = 4$, there will be $\binom{4}{2} = 6$ possible assignment outcomes and the number of points in the sample space would be $6^P$, where $P$ is the number of permuted blocks. Thus, a trial with 100 subjects would have $6^{25} = 2.8 \times 10^{19}$ points in the sample space. It would be impossible to enumerate such a large sample space. However, the randomization distribution may be approximated by using a normal distribution utilizing lower order moments or sampling the sample space.

Define $S_A^j = \sum_{i=1}^{N}\delta_{ij}y_{ij}$ and since $S_A^j$ is a function of $\{\delta_{ij}\}$, we have

$$E\left[S_A^j\bigg|N_A^j = \frac{N}{2}\right] = \frac{1}{2}N\bar{y}_j \quad \text{where } \bar{y}_j = \frac{1}{N}\sum_{i=1}^{N}y_{ij};$$

$$\sigma_j^2 = V\left(S_A^j\bigg|N_A^j = \frac{N}{2}\right) = \frac{N}{4(N-1)}\sum_{i=1}^{N}(y_{ij} - \bar{y}_j)^2 \quad \text{for } j = 1, 2, \ldots, P.$$

Then the analysis proceeds by defining $S_A = \sum_{j=1}^{P} S_A^j$ and

$$E\left[S_A \Big| N_A^j = \frac{N}{2}, j = 1, 2, \ldots, P\right] = \frac{1}{2} N \sum_{j=1}^{P} \bar{y}_j;$$

$$\operatorname{Var}\left(S_A \Big| N_A^j = \frac{N}{2}, j = 1, 2, \ldots, P\right) = \sigma^2 = \sum_{j=1}^{P} \sigma_j^2.$$

Since permuted blocks are independent, by the central limit theorem, as $P$ becomes large, $(S_A - E[S_A | N_A^j = \frac{N}{2}, j = 1, 2, \ldots, P])/\sigma$ will have an approximate standard normal distribution under the hypothesis of no difference between the two treatments; that is,

$$S_A - \tfrac{1}{2} N \sum_{j=1}^{P} \bar{y}_j \stackrel{\cdot}{\sim} \mathcal{N}\left(0, \sum_{j=1}^{P} \sigma_j^2\right).$$

### 3. Multi-center clinical trials.

3.1. *Treatment assignment and covariates.* A characterization of clinical trials is that patients are entered in a trial in a sporadic fashion over time. Ordinarily most clinical trials are designed so that there are approximately equal numbers of patients for each treatment at any point in time in the accrual phase of a trial. This is accomplished by using permuted blocks to randomize patients over time. If there are other prognostic factors influencing the outcomes, in the absence of stratified randomization, the number of patients on each treatment within the level of a prognostic factor would be a random variable. Multi-center trials are the most notable example of this phenomenon. When patients are randomized over time, the number of patients on each treatment, within an institution, is a random variable. Furthermore, if there are other prognostic variables, the number of patients assigned to each treatment at each variable level will be a random variable.

Note that if the trial design uses permuted blocks over time, then one cannot carry out randomization with an institution. Alternatively, a trial may be designed so that the randomization process is stratified by institution. Then permuted blocks over time cannot be implemented. If the institutions enroll a relatively large number of patients, then it will be feasible to utilize permuted blocks within an institution. However, most trials have a large number of institutions which enroll a relatively small number of patients. Consequently, it is not feasible to stratify using permuted blocks within an institution. As noted earlier, for most clinical trials, randomization schemes are guided by permuted blocks over time.

In general, the method of analysis to adjust for a prognostic variable is to condition on the ancillary statistics. The resulting conditional distribution of the test statistic may be impossible to derive because it is necessary to also condition on the permuted blocks. When there are two or more prognostic variables, the conditioning can account for interactions between the prognostic variables. The method of analysis will be illustrated by applying it to account for institutional variation. It should be noted that, in practice, the usual model-based analyses do not account for institutional variation.

3.2. *Multi-center trials.* In our development of multi-center trials, it will be assumed that permuted blocks allocate patients to treatments over time. As a result, the number of patients assigned to each treatment within an institution will be a random variable. These are ancillary statistics and the analysis will condition on patient treatment numbers within each institution.

We will use the following notation:
$P$ = number of permuted blocks;
$K$ = number of institutions;
$N_{jk}$ = number of patients from institution $k$ in the $j$th block;
$N_{\cdot k} = \sum_{j=1}^{P} N_{jk}$ = number of patients from institution $k$;
$n_{kA}^{j}$ = number of patients assigned to $A$ from institution $k$ in the $j$th block;
$n_{kA} = \sum_{j=1}^{P} n_{kA}^{j}$ = number of patients assigned to $A$ in $k$th institution;
$y_{ij}$ = outcome of $i$th patient in $j$th block ($i = 1, 2, \ldots, N$);
$\delta_{ij} = 1$ if $i$th patient in $j$th block is assigned to $A$; 0 otherwise.
It will be convenient to set $Y_j = (y_{1j}, y_{2j}, \ldots, y_{Nj})'$.

In considering the randomization distribution, the quantities $\{N_{jk}\}$ are assumed to be fixed. Without conditioning, enumerating all possible treatment assignments within blocks will result in different $\{n_{kA}^{j}\}$ and $\{n_{kA}\}$ values as the block assignments are independent of institutions. The number of points in the sample space is

$$\mathcal{S}(\{n_{kA}^{j}\}) = \sum_{\mathcal{S}^*} \prod_{j=1}^{P} \prod_{k=1}^{K} \binom{N_{jk}}{n_{kA}^{j}},$$

where we define $\binom{m}{n} = 1$ if either $m$ or $n$ is 0 and the summation is over the set $\mathcal{S}^* = \{\sum_{k=1}^{K} n_{kA}^{j} = \frac{N}{2}, j = 1, 2, \ldots, P\}$.

If we condition on $\{n_{kA}\}$, enumerating the treatment assignments will be restricted to keeping $\{n_{kA}\}$ as constants, that is, $\{n_{kA}^{j}\}$ can result in different numerical values depending on treatment assignment, but is restricted to values that satisfy $\sum_{j=1}^{P} n_{kA}^{j} = n_{kA}$ = constant, for $k = 1, 2, \ldots, K$. The conditional sample space has fewer points and is explicitly

$$\mathcal{T}(\{n_{kA}^{j}\}) = \sum_{\mathcal{T}^*} \prod_{j=1}^{P} \prod_{k=1}^{K} \binom{N_{jk}}{n_{kA}^{j}},$$

where $\mathcal{T}^* = \{\sum_{k=1}^{K} n_{kA}^j = \frac{N}{2}, j = 1,\ldots,P; \sum_{j=1}^{P} n_{kA}^j = n_{kA}, k = 1,\ldots,K\}$.
Consequently, the conditional distribution of the $\{n_{kA}^j\}$ is $\mathcal{T}(\{n_{kA}^j\})/\mathcal{S}(\{n_{kA}^j\})$.
With $K$ more restrictions on the possible assignments across the blocks, the $\{n_{kA}^j\}$ are no longer independent of the blocks. As a result, the mean and variance of the test statistic are difficult to calculate. Furthermore, resampling from the set $\mathcal{T}$ is not readily carried out, so that resampling cannot be used to approximate the distribution.

In order to approximate the conditional distribution of $S_A$, as defined in Section 2.2, based on low order moments, we will consider the randomization process as being generated by a permuted block design. The following notation is needed. Define

$$I_{ijk} = \begin{cases} 1, & \text{if } i\text{th patient in } j\text{th block is from institution } k, \\ 0, & \text{otherwise.} \end{cases}$$

For each patient within a block, define a $K \times 1$ vector $I_{ij} = (I_{ij1}, I_{ij2}, \ldots, I_{ijK})'$, $i = 1,\ldots,N$; $j = 1,\ldots,P$. Consequently, for each block, there exists a $N \times K$ matrix $\mathbf{I_j} = (I'_{1j}, I'_{2j}, \ldots, I'_{Nj})'$, $j = 1, 2, \ldots, P$. The basic relations among the $\{I_{ijk}\}$ are

$$\sum_{k=1}^{K} I_{ijk} = 1; \qquad \sum_{i=1}^{N} I_{ijk} = N_{jk}; \qquad \sum_{j=1}^{P}\sum_{i=1}^{N} I_{ijk} = N_{\cdot k};$$

$$\sum_{i=1}^{N} \delta_{ij} I_{ijk} = n_{kA}^j; \qquad \sum_{k=1}^{K}\sum_{i=1}^{N} \delta_{ij} I_{ijk} = \frac{N}{2}; \qquad \sum_{j=1}^{P}\sum_{i=1}^{N} \delta_{ij} I_{ijk} = n_{kA}.$$

3.3. *Approximation using multivariate normal distribution.* In this section we will extend the results for permuted block designs developed in Section 2.2 to multi-center trials. Consider $n_{kA}^j = \sum_{i=1}^{N} \delta_{ij} I_{ijk}$ ($k = 1, 2, \ldots, K$) and $S_A^j = \sum_{i=1}^{N} \delta_{ij} y_{ij}$. The joint distribution of $S_A^j$ and $\{n_{kA}^j\}$ will be approximately multivariate normal. We can then obtain the distribution of $S_A$ conditional on $\{n_{kA}\}$.

Note that $S_A$ and $\{n_{kA}\}$ are linear functions of $\{\delta_{ij}\}$; that is,

$$S_A = \sum_{j=1}^{P} S_A^j = \sum_{j=1}^{P}\sum_{i=1}^{N} \delta_{ij} y_{ij};$$

$$n_{kA} = \sum_{j=1}^{P} n_{kA}^j = \sum_{j=1}^{P}\sum_{i=1}^{N} \delta_{ij} I_{ijk}, \qquad k = 1,\ldots,K.$$

Then by extending the results in Section 2.2, we have the conditional mean and variance of $S_A^j$ and $\{n_{kA}^j\}$ for any block $j$; that is,

$$E\left[S_A^j \Big| N_A^j = \frac{N}{2}\right] = \frac{1}{2}\sum_{i=1}^{N} y_{ij} \quad \text{and}$$

$$\operatorname{Var}\left(S_A^j \Big| N_A^j = \frac{N}{2}\right) = \frac{1}{4}\frac{N}{N-1}\sum_{i=1}^{N}(y_{ij}-\bar{y}_j)^2,$$

$$E\left[\mathbf{n_A^j}\Big| N_A^j = \frac{N}{2}\right] = \frac{1}{2}\mathbf{N_{j.}} \quad \text{and}$$

$$\operatorname{Var}\left(\mathbf{n_A^j}\Big| N_A^j = \frac{N}{2}\right) = \frac{1}{4}\frac{N}{N-1}\left(\operatorname{Diag}(\mathbf{N_{j.}}) - \frac{\mathbf{N_{j.}N_{j.}'}}{N}\right),$$

$$\operatorname{Cov}\left(S_A^j, n_{kA}^j \Big| N_A^j = \frac{N}{2}\right) = \frac{1}{4}\frac{N}{N-1} Y_j'\left(\mathbf{I} - \frac{\mathbf{J}}{N}\right) I_{jk}, \quad k=1,2,\ldots,K,$$

where $\mathbf{n_A^j} = (n_{1A}^j, n_{2A}^j, \ldots, n_{KA}^j)'$ and $\mathbf{N_{j.}} = (N_{j1}, N_{j2}, \ldots, N_{jK})'$ and $I_{jk} = (I_{1jk}, I_{2jk}, \ldots, I_{Njk})'$ is the $k$th column in $\mathbf{I_j}$.

Since the permuted blocks are independent, by the multivariate central limit theorem, $S_A$ and $\{n_{kA}\}$ will have an approximate $(K+1)$-multivariate normal distribution, that is,

$$\frac{1}{\sqrt{P}}\left(\begin{pmatrix} S_A \\ \mathbf{n_{.A}} \end{pmatrix} - \frac{1}{2}\begin{pmatrix} \sum_{j=1}^{P}\sum_{i=1}^{N} y_{ij} \\ \mathbf{N_{..}} \end{pmatrix}\right) \dot{\sim} \mathcal{N}\left(\mathbf{0}, \frac{\sum_{j=1}^{P}\mathbf{V_j}}{P}\right),$$

where $\mathbf{n_{.A}} = (n_{1A}, n_{2A}, \ldots, n_{KA})', \mathbf{N_{..}} = (N_{.1}, N_{.2}, \ldots, N_{.K})'$ and

$$\mathbf{V_j} = \begin{pmatrix} \operatorname{Var}\left(S_A^j\Big|N_A^j=\frac{N}{2}\right) & \operatorname{Cov}\left(S_A^j, \mathbf{n_A^j}\Big|N_A^j=\frac{N}{2}\right)' \\ \operatorname{Cov}\left(S_A^j, \mathbf{n_A^j}\Big|N_A^j=\frac{N}{2}\right) & \operatorname{Var}\left(\mathbf{n_A^j}\Big|N_A^j=\frac{N}{2}\right) \end{pmatrix}.$$

Consequently, the conditional distribution of $S_A|\mathbf{n_{.A}}$ will be approximately normal with mean and variance given by standard multivariate normal theory; for example,

$$E[S_A|\{n_{kA}\}] = \frac{1}{2}\sum_{j=1}^{P}\sum_{i=1}^{N} y_{ij} + \left\{\sum_{j=1}^{P}\operatorname{Cov}\left(S_A^j, \mathbf{n_A^j}\Big|N_A^j=\frac{N}{2}\right)'\right\}$$

$$\times \left\{\sum_{j=1}^{P}\operatorname{Var}\left(\mathbf{n_A^j}\Big|N_A^j=\frac{N}{2}\right)\right\}^{-}\cdot\left(\mathbf{n_{.A}} - \frac{1}{2}\mathbf{N_{..}}\right);$$

$$\operatorname{Var}(S_A|\{n_{kA}\}) = \sum_{j=1}^{P}\operatorname{Var}\left(S_A^j\Big|N_A^j=\frac{N}{2}\right) - \left\{\sum_{j=1}^{P}\operatorname{Cov}\left(S_A^j, \mathbf{n_A^j}\Big|N_A^j=\frac{N}{2}\right)'\right\}$$

$$\times \left\{\sum_{j=1}^{P}\operatorname{Var}\left(\mathbf{n_A^j}\Big|N_A^j=\frac{N}{2}\right)\right\}^{-}$$

$$\times \left\{ \sum_{j=1}^{P} \text{Cov}\left( S_A^j, \mathbf{n_A^j} \Big| N_A^j = \frac{N}{2} \right) \right\},$$

where $(\cdot)^-$ is a generalized inverse.

See Harville (1997) for more theory and application of the generalized inverse. A generalized inverse of a matrix can be obtained using most standard statistical software. In the simulation studies (Section 3.4), the **R** [R Development Core Team (2007)] function *ginv* in package MASS was used to generate the Moore–Penrose generalized inverse.

3.4. *Simulation studies.* Simulation studies are carried out to characterize the behavior of the conditional test for continuous, binary and censored outcomes. The simulations also explore the effect of block size and number of institutions. Block sizes are chosen to be 4, 6 and 8. The number of institutions varies from 10 to 140 depending on sample sizes (120, 240 and 360). For each sample size, the average number of patients per institution ranges from 3 to 20. The size of the treatment effect is chosen such that the power of the tests is at meaningful levels for comparison purposes. Continuous outcomes are drawn from a lognormal distribution with treatment difference of 1.07. Institution effects are drawn from a normal distribution with a standard deviation of 2. For binary outcomes, the underlying distribution is Bernoulli with success probabilities 0.5 and 0.7, respectively. A logistic model is used for the treatment and institution effects [drawn from a N(0, 1.73)]. In the case of censored outcomes, the survival times are generated from exponential distributions with a ratio of 1.5 for the two treatment mean survival times. Institution effects are introduced using a multiplicative model and are drawn from a $\chi^2$ distribution with 1 df.

When the outcomes are right-censored time-to-event data with noninformative censoring, the outcome variable $y_{ij}$ is replaced with specific test statistics calculated within each block and then aggregated over the blocks. In practice, the logrank score or Gehan score is used to test the null hypothesis that there is no difference in the survival of the individuals in the two groups. To test the null hypothesis that the mortality rate is the same for the two groups, mortality rates are calculated and used in place of $\bar{y}_j$. We used both logrank and Gehan scores in the simulation studies. The simulations were set up so that each group had the same censoring rate.

For all three types of outcomes, the power of the conditional test was compared with commonly used tests ignoring institutions. Comparisons were made both ignoring blocks and stratifying by blocks. For continuous outcomes, the conditional test was compared with the two-sample $t$ test based on the treatment contrast within each block, which we denote as a "stratified $t$ test" in the tables. Also, comparisons were made for a two-sample $t$ test,

TABLE 1
*Power comparison for continuous outcomes with no block variation*

| Sample size No. of institutions | 120 | | | 240 | | | | 360 | | | | |
|---|---|---|---|---|---|---|---|---|---|---|---|---|
| | 10 | 20 | 40 | 20 | 40 | 60 | 80 | 20 | 40 | 60 | 80 | 100 |
| Block size 4 | | | | | | | | | | | | |
| Conditional test | 0.53 | 0.47 | 0.35 | 0.81 | 0.76 | 0.69 | 0.62 | 0.93 | 0.92 | 0.90 | 0.85 | 0.82 |
| Randomization test | 0.42 | 0.40 | 0.39 | 0.68 | 0.70 | 0.67 | 0.67 | 0.85 | 0.84 | 0.84 | 0.83 | 0.85 |
| Stratified t | 0.43 | 0.40 | 0.40 | 0.68 | 0.68 | 0.67 | 0.67 | 0.85 | 0.84 | 0.84 | 0.83 | 0.85 |
| t-test | 0.43 | 0.40 | 0.40 | 0.68 | 0.68 | 0.67 | 0.67 | 0.85 | 0.84 | 0.84 | 0.83 | 0.85 |
| Block size 8 | | | | | | | | | | | | |
| Conditional test | 0.53 | 0.48 | 0.39 | 0.81 | 0.76 | 0.72 | 0.65 | 0.94 | 0.93 | 0.91 | 0.89 | 0.85 |
| Randomization test | 0.42 | 0.40 | 0.39 | 0.69 | 0.67 | 0.68 | 0.66 | 0.85 | 0.84 | 0.84 | 0.84 | 0.84 |
| Stratified t | 0.42 | 0.40 | 0.39 | 0.69 | 0.67 | 0.68 | 0.66 | 0.85 | 0.84 | 0.84 | 0.84 | 0.84 |
| t-test | 0.42 | 0.40 | 0.40 | 0.69 | 0.67 | 0.68 | 0.67 | 0.85 | 0.84 | 0.84 | 0.84 | 0.84 |

NOTE. Data are generated from a lognormal distribution. The difference between the two group means is 1.07.

which ignores both institutions and blocks. In the case of binary outcomes, the conditional test was compared with both the Mantel–Haenszel test of $P$ $2 \times 2$ tables with each block forming a $2 \times 2$ table and a simple $2 \times 2$ table which pooled all the data. For censored outcomes, the reference tests are the block stratified Gehan and logrank tests and the logrank test ignoring blocks and institutions. Those reference tests are chosen because, in practice, they are most likely to be employed in the analyses of clinical trials using permuted blocks or ignoring the permuted blocks. Also included in the comparisons was the test based on the randomization distribution generated by permuted blocks, ignoring the conditioning on institutions. The latter test is denoted as "the randomization test" in the tables. Type I error rates were examined where there is no treatment difference. The simulations (results not presented here) showed that all tests had the correct type I error rates at levels 0.01 and 0.05. All censoring was noninformative for censored outcomes.

Tables 1–4 show selected results from the simulations. Each table entry is based on the average of 5000 replications. All tests are two-sided with 0.05 type I error rate. Note that the conditional tests have higher power than their counterparts for each of the three types of outcomes for nearly all institution sizes when institution variation is present. The power of the conditional test decreases as the number of institutions increases for a fixed sample size. This is in contrast to the power of the reference tests which decrease slightly or remain the same for a fixed sample size. Keeping the total sample size fixed, but allowing the number of institutions to increase, results in decreasing the average sample sizes within an institution. Also,

there are more opportunities for treatment assignment imbalances as the number of institutions increases. Treatment imbalances result in decreased statistical efficiency. The imbalances are particularly acute with small institution sample size. For example, an entry of one patient or two patients assigned to the same treatment within an institution carry no information for comparing two treatments.

Tables 1 and 2 show simulation results for continuous outcomes with and without block effects. A block effect is added to the outcomes ($-1$, $-0.5$, $0.5$

TABLE 2
*Power comparison for continuous outcomes with block variation*

| Sample size No. of institutions | 120 | | | 240 | | | | 360 | | | | |
|---|---|---|---|---|---|---|---|---|---|---|---|---|
| | 10 | 20 | 40 | 20 | 40 | 60 | 80 | 20 | 40 | 60 | 80 | 100 |
| Block size 4 | | | | | | | | | | | | |
| Conditional test | 0.52 | 0.47 | 0.34 | 0.81 | 0.75 | 0.69 | 0.59 | 0.93 | 0.92 | 0.89 | 0.86 | 0.81 |
| Randomization test | 0.42 | 0.40 | 0.38 | 0.70 | 0.67 | 0.68 | 0.67 | 0.85 | 0.84 | 0.84 | 0.84 | 0.83 |
| Stratified t | 0.42 | 0.41 | 0.38 | 0.70 | 0.67 | 0.69 | 0.67 | 0.85 | 0.84 | 0.84 | 0.84 | 0.83 |
| t-test | 0.39 | 0.38 | 0.37 | 0.68 | 0.65 | 0.67 | 0.65 | 0.84 | 0.83 | 0.83 | 0.82 | 0.82 |
| Block size 8 | | | | | | | | | | | | |
| Conditional test | 0.54 | 0.48 | 0.38 | 0.82 | 0.77 | 0.72 | 0.64 | 0.94 | 0.92 | 0.90 | 0.88 | 0.84 |
| Randomization test | 0.41 | 0.40 | 0.40 | 0.69 | 0.69 | 0.67 | 0.67 | 0.84 | 0.84 | 0.84 | 0.84 | 0.84 |
| Stratified t | 0.42 | 0.40 | 0.40 | 0.69 | 0.69 | 0.67 | 0.68 | 0.84 | 0.84 | 0.84 | 0.84 | 0.84 |
| t-test | 0.39 | 0.37 | 0.37 | 0.67 | 0.67 | 0.66 | 0.65 | 0.83 | 0.83 | 0.82 | 0.83 | 0.82 |

NOTE. Data are generated from a lognormal distribution. The difference between the two group means is 1.07.

TABLE 3
*Power comparison for binary outcomes*

| Sample size No. of institutions | 120 | | | 240 | | | | 360 | | | | |
|---|---|---|---|---|---|---|---|---|---|---|---|---|
| | 10 | 20 | 40 | 20 | 40 | 60 | 80 | 20 | 40 | 60 | 80 | 100 |
| Block size 4 | | | | | | | | | | | | |
| Conditional test | 0.37 | 0.32 | 0.25 | 0.64 | 0.58 | 0.52 | 0.46 | 0.82 | 0.79 | 0.76 | 0.72 | 0.67 |
| Mantel–Haenszel & RT[a] | 0.30 | 0.27 | 0.27 | 0.48 | 0.47 | 0.49 | 0.49 | 0.65 | 0.65 | 0.65 | 0.65 | 0.65 |
| Single $2 \times 2$ table | 0.30 | 0.28 | 0.28 | 0.49 | 0.48 | 0.50 | 0.50 | 0.66 | 0.65 | 0.66 | 0.65 | 0.65 |
| Block size 8 | | | | | | | | | | | | |
| Conditional test | 0.45 | 0.41 | 0.32 | 0.72 | 0.68 | 0.63 | 0.58 | 0.89 | 0.87 | 0.84 | 0.82 | 0.79 |
| Mantel–Haenszel & RT | 0.39 | 0.38 | 0.37 | 0.63 | 0.62 | 0.62 | 0.63 | 0.80 | 0.80 | 0.79 | 0.80 | 0.81 |
| Single $2 \times 2$ table | 0.40 | 0.39 | 0.38 | 0.64 | 0.63 | 0.64 | 0.65 | 0.80 | 0.80 | 0.80 | 0.80 | 0.81 |

NOTE. Data are generated from the Bernoulli distribution. Success probabilities for two treatments are 0.5 and 0.7 respectively.
[a]RT refers to the randomization test ignoring institutions.

and 1 each for one quarter of the blocks) to reflect a trend. With block effects, the superiority of the conditional test tends to be greater compared to the t-test. The randomization test and the stratified t-test also outperform the t-test as their performances are not affected by the block variation. Also, it is more noticeable that, with block variation, the power of the conditional tests is slightly higher with larger block sizes.

**4. Conditional group sequential tests based on randomization.** Jennison and Turnbull (1997) have extended group sequential analysis to incorporate covariates in a wide range of generalized linear models and the proportional hazards model for survival data. However, to our knowledge, there is no method developed for group sequential analysis using randomization based tests which also adjusts for covariates.

We illustrate the details of the conditional group sequential test for continuous outcomes. Procedures for censored outcomes and proportions are similar. Suppose we intend to monitor the data to a maximum of $L$ interim analyses. At the $l$th interim analysis, let $P_l$ refer to the cumulative number of permuted blocks utilized up to this time. Hence, the number of patients at the $l$th interim analysis will be $NP_l$, where $N$ is the block size. Let the information fraction $t_l = P_l/P_{\max}$, where $P_L = P_{\max}$. The test statistic $S_A(t_l)$

TABLE 4
*Power comparison for censored outcomes*

| Sample size No. of institutions | 120 | | | 240 | | | | 360 | | | | |
|---|---|---|---|---|---|---|---|---|---|---|---|---|
| | 10 | 20 | 40 | 20 | 40 | 60 | 80 | 20 | 40 | 60 | 80 | 100 |
| Block size 4 | | | | | | | | | | | | |
| CT (Gehan score)[a] | 0.23 | 0.19 | 0.14 | 0.39 | 0.34 | 0.29 | 0.25 | 0.55 | 0.50 | 0.46 | 0.41 | 0.39 |
| Stratified Gehan | 0.16 | 0.14 | 0.12 | 0.24 | 0.23 | 0.21 | 0.22 | 0.34 | 0.31 | 0.30 | 0.29 | 0.29 |
| CT (Logrank score)[a] | 0.23 | 0.18 | 0.14 | 0.38 | 0.34 | 0.27 | 0.25 | 0.54 | 0.49 | 0.44 | 0.40 | 0.38 |
| RT (Logrank score)[a] | 0.16 | 0.15 | 0.13 | 0.25 | 0.23 | 0.21 | 0.22 | 0.36 | 0.32 | 0.31 | 0.30 | 0.30 |
| Stratified Logrank | 0.16 | 0.15 | 0.13 | 0.25 | 0.23 | 0.21 | 0.22 | 0.36 | 0.33 | 0.31 | 0.30 | 0.31 |
| Logrank test | 0.21 | 0.19 | 0.18 | 0.33 | 0.31 | 0.30 | 0.30 | 0.46 | 0.43 | 0.42 | 0.42 | 0.41 |
| Block size 8 | | | | | | | | | | | | |
| CT (Gehan score)[a] | 0.27 | 0.25 | 0.20 | 0.47 | 0.41 | 0.37 | 0.33 | 0.64 | 0.59 | 0.56 | 0.53 | 0.49 |
| Stratified Gehan | 0.17 | 0.16 | 0.14 | 0.26 | 0.24 | 0.22 | 0.22 | 0.36 | 0.34 | 0.32 | 0.33 | 0.32 |
| CT (Logrank score)[a] | 0.27 | 0.24 | 0.19 | 0.45 | 0.39 | 0.35 | 0.31 | 0.62 | 0.57 | 0.53 | 0.51 | 0.47 |
| RT (Logrank score)[a] | 0.18 | 0.16 | 0.16 | 0.27 | 0.25 | 0.25 | 0.24 | 0.39 | 0.36 | 0.34 | 0.35 | 0.36 |
| Stratified Logrank | 0.18 | 0.16 | 0.16 | 0.27 | 0.26 | 0.25 | 0.24 | 0.39 | 0.36 | 0.35 | 0.35 | 0.36 |
| Logrank test | 0.21 | 0.19 | 0.18 | 0.31 | 0.31 | 0.30 | 0.29 | 0.45 | 0.43 | 0.41 | 0.42 | 0.42 |

NOTE. Data are generated from an exponential distribution. Ratio of two mean survival times is 1.5. Percentage of censored observations is 19%.
[a]CT refers to the conditional test using the Gehan or the logrank score; RT refers to the randomization test ignoring institutions using the logrank score.

is the sum of the inner product of the $\{\delta_{ij}\}$ and outcomes $\{y_{ij}\}$ in each block up to the $P_l$th block, that is,

$$S_A(t_l) = \sum_{j=1}^{P_l} S_A^j = \sum_{j=1}^{P_l} \sum_{i=1}^{N} \delta_{ij} y_{ij}.$$

Conditional on $\mathbf{n_{.A}(t_l)} = (n_{1A}(t_l), n_{2A}(t_l), \ldots, n_{KA}(t_l))'$, where $n_{kA}(t_l)$ is the number of patients assigned to treatment $A$ in the $k$th institution up to the time before the $l$th interim analysis, $S_A(t_l)$ is approximately normal with mean and variance given in Section 3.3.

The test statistic $T(t_l)$ at each interim analysis is computed as

$$T(t_l) = \frac{S_A(t_l) - E[S_A(t_l)|\mathbf{n_{.A}(t_l)}]}{\sqrt{\mathrm{Var}(S_A(t_l)|\mathbf{n_{.A}(t_l)})}}$$

and compared to a pre-specified boundary $C_l$, $l = 1, 2, \ldots, L$. The null hypothesis of no difference in treatments is rejected if $|T(t_l)| > C_l$ for some $l = 1, \ldots, L$.

We carried out simulations comparing conditional group sequential tests and their counterparts for continuous, binary and censored outcomes. The simulation setting was chosen to be the same for all types of outcomes. The test was one-sided with significance level 0.025. Equal numbers of patients (240) coming from a number of institutions (ranging from 10 to 60) are assigned to the two treatments using a permuted block design (Block sizes: 4, 6 and 8). Treatment differences and institution effects followed the same data generating scheme as in section 3.4. For continuous outcome, the institution effects were drawn from a N(0, 2) and for binary outcome, a logistic model was used and institution effects were drawn from a N(0, 1.73). For censored outcomes, institution effects were drawn from a scaled $\chi^2$ distribution with 4 df. Block effects were not implemented in the simulation. Four interim analyses ($L = 4$) were made where the increment of information changed by $\frac{1}{4}$. Stopping boundaries were calculated using the commonly used O'Brien–Fleming rule [O'Brien and Fleming (1979)], which is a special case of the $\alpha$ spending approach [Lan and DeMets (1983)]. The O'Brien–Fleming boundary gives $C_l = 2.024(4/l)^{1/2}$ for $l = 1, 2, 3$ and 4. In our simulations we first checked the type I error rate and found that the observed size of the test was satisfactory close to the nominal level. Table 4 summarizes a comparison of the conditional group sequential tests verses the usual group sequential tests (unconditional) for block sizes of $N = 4$ and 8 for continuous, binary and censored outcomes. For all three types of outcomes, the conditional tests have higher power than the unconditioned tests; cf. Table 5.

**5. Discussion.** In this paper we investigated methods of analysis which are guided by the design of the study. More specifically, the analysis is based on the randomization process generated by permuted blocks that are used to allocate treatments to patients over time. Another feature of this paper is to take account of the effects associated with hospitals/treatment centers.

Institutional variation abounds in nearly all clinical trials. The more severe/common the side effects, the greater the reliance on patient management in a center. Discontinuation of treatment due to not managing the side effects properly will result in different institutional outcomes in any intent to treat analysis. Since many trials may be carried out with both community hospitals and major treatment centers, there is substantial variation among hospitals with regard to patient management. Trials involving surgery, as an adjunct treatment, show institution preference for different surgical procedures. Patient recruitment is also an important factor leading to institutional differences among populations. In some institutions, patients with co-morbid disease are not approached to enter a trial despite being eligible. The reporting of patient refusals to participate in a clinical trial is not usually done in a published paper. However, there are large variations in the declination rate among institutions, which reflect on the different populations being entered on a trial. Closely related is the proportion of eligible patients actually entered in a trial. Intensive safety monitoring generally leads to more favorable outcomes. Such monitoring greatly varies within centers. There may be important institutional differences in the use of ancillary nontrial treatments. For example, trials of aspirin and heparin in patients with acute ischaemic stroke often receive

TABLE 5
*Power comparison in the presence of institutional variation for group sequential tests: conditional group sequential tests vs. standard tests (group sequential tests without conditioning)*

| Outcomes | Type of test | Block size 4 No. of institutions | | | | Block size 8 No. of institutions | | | |
|---|---|---|---|---|---|---|---|---|---|
| | | 10 | 20 | 40 | 60 | 10 | 20 | 40 | 60 |
| Continuous | Conditional | 0.72 | 0.70 | 0.69 | 0.67 | 0.72 | 0.73 | 0.70 | 0.68 |
| | Unconditional | 0.51 | 0.49 | 0.48 | 0.47 | 0.52 | 0.50 | 0.46 | 0.48 |
| Binary | Conditional | 0.91 | 0.91 | 0.90 | 0.88 | 0.95 | 0.96 | 0.95 | 0.94 |
| | Unconditional | 0.76 | 0.76 | 0.77 | 0.77 | 0.89 | 0.89 | 0.89 | 0.90 |
| Censored[a] | Conditional (Logrank score) | 0.64 | 0.63 | 0.62 | 0.56 | 0.86 | 0.83 | 0.83 | 0.80 |
| | Stratified Logrank | 0.57 | 0.57 | 0.57 | 0.57 | 0.78 | 0.77 | 0.75 | 0.75 |

NOTE. Sample size is 480 with 4 looks at equal information increments. One-sided test with $\alpha = 0.025$.
[a]Percentage of censored outcomes is 18.5%.

nontrial treatments such as glycerol and steroids. The European Carotid Surgery Trial on endarterectomy for symptomatic carotid stenosis showed substantial differences in the speed with which patients were entered in the trial based on the last symptomatic event. It ranged from weeks to months depending on institutions. Benefit from endarterectomy depended significantly on delay to surgery after the presenting event [Rothwell et al. (2004)]. Cancer chemotherapy trials are carried out among major cancer centers as well as community hospitals. The differences among centers with regard to patient management are large—especially in the management of side effects. As clinical trials become more global, there may be substantial differences within a country on methods of diagnosis and treatment, probably resulting in even more variation among centers. The effects of the centers may be ameliorated by conditioning on the ancillary statistics which are the treatment sample sizes within each institution. In practice, the institution variation can be substantial. For example, the trial used as an illustration by Skene and Wakefield (1990) showed that the variation of placebo response rates was of the same magnitute as the variation of the treatment effect. In another example, Yamaguchi and Ohashi (1999) reported a much larger center variation for the baseline risk than the variation of the treatment effects in a multi-center clinical bladder cancer trial. Andersen, Klein, and Zhang (1999) showed that if center effect is ignored, the estimator of the main treatment effect may be quite biased and is inconsistent.

We have shown that the design based analysis, in the presence of institution variation, results in greater power for commonly used clinical trial designs, compared to model based analyses in which institutions, permuted blocks and the randomization process are ignored. This is true for binary, continuous and censored outcomes. Also, this advantage holds for group sequential trials.

The novel idea of conditioning on the ancillary statistics provides an alternative method to adjust for covariates in a randomization based inference. Conditioning on the sample sizes of institutions is an illustration of how a randomization based analysis of a trial may adjust for covariates. In this case the centers are the covariates and the sample size of each center is an ancillary statistic. For an arbitrary covariate, the number of patients assigned to one treatment for each level of the covariate is an ancillary statistic. Conditioning on the ancillary statistics in a randomization based analysis is a way of adjusting for the covariate effect. This idea generalizes when there are an arbitrary number of covariates. Discretized continuous covariates can also be adjusted using the same idea. Another approach to adjust for the continuous covariates is to employ a regression model and use the residuals from the model as responses in the proposed methods. Details of this approach will be discussed in a follow-up paper.

The conditional test is based on the randomization distribution generated by the random assignment of treatments to patients. It does not make any assumptions about sampling from a target population. Thus, inference based on the randomization distribution is applicable to patients who entered the study. Alternatively, model based analyses assume that patients are random samples drawn from a population and that centers are also randomly drawn from a population of centers. This assumption is rarely true. In fact, patients are recruited into a study from a nonrandom selection of centers. Centers are often chosen because of their affiliations, locations and/or expertise. Within each center, patients are recruited because they are eligible and willing to participate in randomized clinical trials. Thus, neither the centers nor the patients are random samples. This raises the issue of whether the assumptions of the model based analysis are correct. Lachin (1988) concluded that using model based analysis on a randomized clinical trial "becomes a matter of faith that is based upon assumptions that are inherently untestable." Ludbrook and Dudley (1998) surveyed 252 comparative studies published in 5 biomedical journals and concluded that randomization tests are superior to $t$ and $F$ tests in biomedical research where randomization is the norm rather than random sampling from patient and institution populations.

One of the reviewers has asked about whether randomization techniques can be used for estimation in the context of particular models. Confidence intervals can be obtained by inverting the test. Garthwaite (1996) used simulation in conjunction with a Robbins–Monro search process to locate the two ends of the confidence interval by inverting randomization tests. For censored observation, confidence intervals from the inversion of normal tests are described in Kalbfleisch and Prentice (2002). Alternatively, one can also obtain confidence intervals by simulation. For example, if the ratio of two mortalities is the parameter of interest, let $\hat{p}$ be the observed ratio of mortalities (m1/m2, where m = total number of deaths/total followup time). Order the survival times from the smallest to the largest. At each time when there is a death, assign the observation to treatment 1 with probability $\hat{p} * n_1/(\hat{p} * n_1 + n_2)$, where $n_i$ is the number of people at risk for treatment $i$. At each censored time, assign the observation to treatment 1 with probability $n_1/(n_1 + n_2)$. The ratio of mortalities can be calculated for each realization. One can approximate the distribution of the ratio of mortalities by repeating the rerandomization a large number of times. The 2.5th and 97.5th percentile values of the distribution are the 95% confidence limits for the ratio of two mortalities. Initial numeric studies have shown correct coverage probabilities of the confidence intervals under various settings including the null and alternative hypothesis, when the true mortality ratio is less than 2. We have found that the coverage probabilities may tend to be less than 0.95 as the true ratio of mortalities tend to be greater than 2. In practice, a mortality ratio of 2 or greater is unlikely to be encountered in a trial. Such a superiority is likely to be shown in pilot and preliminary studies. It may be unethical to evaluate such a large outcome discrepency in a clinical trial.

In conclusion, we advocate that analyses of randomized multi-center clinical trials should be guided by the design of the trial and rely on the randomization process for making the inference. The methods discussed in this paper only rely on the randomization process. Consequently, they are distribution free and are capable of accounting for institutional variation and time trends in patient populations. In practice, the methods discussed in this paper may lead to greater power than the conventional analysis, when there are institution and time trends (block effects). When there is an absence of institution and/or block effects, the power of the conditional test will be lower than statistical tests which ignore these effects. This possibility may be regarded as the "insurance" one must pay in accounting for potential institution and block effects.

The conditional tests discussed here may be generalized to deal with more complex situations. The generalizations can incorporate more than two treatments, stratified permuted block designs, the post-hoc modeling of covariates and missing observations. In a follow-up paper we intend to discuss these topics.

**Acknowledgments.** We thank the associate editor and a referee for constructive comments.

Department of Biostatistics
Harvard School of Public Health
655 Huntington Avenue
Boston, Massachusetts 02115
USA
E-mail: [lzheng@hsph.harvard.edu](lzheng@hsph.harvard.edu)

Department of Biostatistics
Harvard School of Public Health and
Dana-Farber Cancer Institute
44 Binney Street
Boston, Massachusetts 02115
USA
E-mail: [zelen@hsph.harvard.edu](zelen@hsph.harvard.edu)